\documentclass[12pt,a4]{article}
\usepackage{amsmath}
\usepackage{bm}
\usepackage{amsfonts}
\usepackage{amssymb}
\usepackage{graphics}
\usepackage[normal]{caption2}
\usepackage{subfigure}
\usepackage{rotating}
\usepackage{citesort}
\def\be{\begin{equation}}
\def\ee{\end{equation}}
\def\ba{\begin{array}}
\def\ea{\end{array}}
\def\bea{\begin{eqnarray}}
\def\eea{\end{eqnarray}}

\begin{document}
\baselineskip 20pt \setlength\tabcolsep{2.5mm}
\renewcommand\arraystretch{1.5}
\setlength{\abovecaptionskip}{0.1cm}
\setlength{\belowcaptionskip}{0.5cm}
\pagestyle{empty}
\newpage
\pagestyle{plain} \setcounter{page}{1} \setcounter{lofdepth}{2}
\begin{center} {\large\bf Impact parameter dependence of the isospin effects and mass dependence of balance energy}\\
\vspace*{0.4cm}
{\bf Sakshi Gautam}\footnote{Email:~ sakshigautm@gmail.com}\\
{\it  Department of Physics, Panjab University, Chandigarh -160
014, India.\\}
\end{center}
We study the effect of isospin degree of freedom on the balance
energy (E$_{bal}$) as well as its mass dependence throughout the
mass range for two different sets of isobaric systems with N/Z = 1
and 1.4 at different colliding geometries ranging from the central
to peripheral ones. Our findings reveal the dominance of Coulomb
repulsion in isospin effects on E$_{bal}$ as well as its mass
dependence throughout the range of the colliding geometry. Our
results also indicate that the effect of symmetry energy on the
energy of vanishing flow is uniform throughout the mass range and
throughout the colliding geometry.

\newpage
\baselineskip 20pt
\section{Introduction}

The existing and upcoming radioactive ion beam (RIB) facilities
\cite{rib1,rib2,rib3,rib5} around the world led to the increased
interest in isospin degree of freedom and equation of asymmetric
nuclear matter. The above study helps to isolate the
isospin-dependent part of nuclear equation of state which is vital
in understanding the astrophysical phenomena. Isospin degree of
freedom plays an important role through both the nuclear matter
equation of state and as well as via in-medium cross section.
Therefore, heavy-ion collisions induced by neutron-rich nuclei
provide a good opportunity to explore the isospin dependence of
in-medium nuclear interactions.
\par
 After about three decades of intensive efforts in both
nuclear experiments and theoretical calculations, equation of
state for isospin symmetric matter is now relatively well
understood. The effect of isospin degree of freedom on the
collective transverse in-plane flow as well as on its
disappearance \cite{sche68,bon87,krof89} (there exists a
particular incident energy called \emph{balance energy}
(E$_{bal}$) or \emph{energy of vanishing flow} (EVF) at which
transverse in-plane flow disappears) has been reported in the
literature \cite{li96,pak97,daff98}, where it was found
 that neutron-rich systems have higher  E$_{bal}$ compared to neutron-deficient
  systems at all colliding geometries varying from central to peripheral
ones. The effect of isospin degree of freedom on  E$_{bal}$ was
found to be much more pronounced at peripheral colliding
geometries
 compared to central ones. As reported in the literature, the isospin
dependence of collective flow as well as its disappearance has
been explained as a competition among various reaction mechanisms,
such as nucleon-nucleon collisions, symmetry energy, surface
property of the colliding nuclei, and Coulomb force. The relative
importance among these mechanisms is not yet clear \cite{li96}. In
recent study, we \cite{gaum10} confronted theoretical calculations
(using isospin-dependent quantum molecular dynamics (IQMD) model
\cite{hart98}) with the data at all colliding geometries and were
able to reproduce the data within 5\% on the average at all
colliding geometries. Motivated by this good agreement of the
calculations with data, we \cite{gaum10} studied the isospin
effects on the E$_{bal}$ throughout the mass range 48-350 for two
sets of isobaric systems with N/Z = 1.0 and 1.4 at semi central
colliding geometry. These results showed that the difference
between the E$_{bal}$ for two isobaric
 systems is mainly due to the Coulomb repulsion. It was also shown that
  Coulomb repulsion dominates over symmetry energy.
These findings also indicated towards the dominance of the Coulomb
 repulsion in larger magnitude of isospin effects in
 E$_{bal}$ at peripheral collisions. Here we aim to extend the study over
  full range of colliding geometry varying from central to
 peripheral ones. Section 2
describes the model in brief. Section 3 explains the results and
discussion and Sec. 4 summarizes the results.
\section{The model}
The present study is carried out within the framework of the IQMD
model \cite{hart98} which is an extension of the QMD model
\cite{puri1,puri2,puri3}. The IQMD model treats different charge
states of nucleons, deltas, and pions explicitly, as inherited
from the Vlasov-Uehling-Uhlenbeck (VUU) model. The IQMD model has
been used successfully for the analysis of a large number of
observables from low to relativistic energies. The isospin degree
of freedom enters into the calculations via symmetry potential,
cross sections, and Coulomb interaction.
 \par
 In this model, baryons are represented by Gaussian-shaped density distributions
\begin{eqnarray}
f_{i}(\vec{r},\vec{p},t) =
\frac{1}{\pi^{2}\hbar^{2}}\exp(-[\vec{r}-\vec{r_{i}}(t)]^{2}\frac{1}{2L})
\nonumber\\~~~
 \times\exp(-[\vec{p}- \vec{p_{i}}(t)]^{2}\frac{2L}{\hbar^{2}})
 \end{eqnarray}
 Nucleons are initialized in a sphere with radius R = 1.12 A$^{1/3}$ fm, in accordance with liquid-drop model.
 Each nucleon occupies a volume of \emph{h$^{3}$}, so that phase space is uniformly filled.
 The initial momenta are randomly chosen between 0 and Fermi momentum ($\vec{p}$$_{F}$).
 The nucleons of the target and projectile interact by two- and three-body Skyrme forces, Yukawa potential, Coulomb interactions,
  and momentum-dependent interactions. In addition to the use of explicit charge states of all baryons and mesons, a symmetry potential between protons and neutrons
 corresponding to the Bethe-Weizsacker mass formula has been included. The hadrons propagate using Hamilton equations of motion:
\begin {eqnarray}
\frac{d\vec{{r_{i}}}}{dt} = \frac{d\langle H
\rangle}{d\vec{p_{i}}};& & \frac{d\vec{p_{i}}}{dt} = -
\frac{d\langle H \rangle}{d\vec{r_{i}}}
\end {eqnarray}
 with
\begin {eqnarray}
\langle H\rangle& =&\langle T\rangle+\langle V \rangle
\nonumber\\
& =& \sum_{i}\frac{p^{2}_{i}}{2m_{i}} + \sum_{i}\sum_{j>i}\int
f_{i}(\vec{r},\vec{p},t)V^{ij}(\vec{r}~',\vec{r})
 \nonumber\\
& & \times f_{j}(\vec{r}~',\vec{p}~',t) d\vec{r}~ d\vec{r}~'~
d\vec{p}~ d\vec{p}~'.
\end {eqnarray}
 The baryon potential\emph{ V$^{ij}$}, in the above relation, reads as
 \begin {eqnarray}
  \nonumber V^{ij}(\vec{r}~'-\vec{r})& =&V^{ij}_{Sky} + V^{ij}_{Yuk} +
  V^{ij}_{Coul} + V^{ij}_{mdi} + V^{ij}_{sym}
    \nonumber\\
   & =& [t_{1}\delta(\vec{r}~'-\vec{r})+t_{2}\delta(\vec{r}~'-\vec{r})\rho^{\gamma-1}(\frac{\vec{r}~'+\vec{r}}{2})]
   \nonumber\\
   &  & +t_{3}\frac{\exp(|(\vec{r}~'-\vec{r})|/\mu)}{(|(\vec{r}~'-\vec{r})|/\mu)}+
    \frac{Z_{i}Z_{j}e^{2}}{|(\vec{r}~'-\vec{r})|}
   \nonumber \\
   &  & +t_{4}\ln^{2}[t_{5}(\vec{p}~'-\vec{p})^{2} +
    1]\delta(\vec{r}~'-\vec{r})
    \nonumber\\
   &  & +t_{6}\frac{1}{\varrho_{0}}T_{3i}T_{3j}\delta(\vec{r_{i}}~'-\vec{r_{j}}).
 \end {eqnarray}
Here \emph{Z$_{i}$} and \emph{Z$_{j}$} denote the charges of
\emph{ith} and \emph{jth} baryon, and \emph{T$_{3i}$} and
\emph{T$_{3j}$} are their respective \emph{T$_{3}$} components
(i.e., $1/2$ for protons and $-1/2$ for neutrons). The
parameters\emph{ $\mu$} and \emph{t$_{1}$,....,t$_{6}$} are
adjusted to the real part of the nucleonic optical potential.
 For the density dependence of  the nucleon optical potential, standard Skyrme type parametrization is
 employed. Note that such Skyrme forces and related potential are
 essential tool to understand low energy phenomena such as fusion,
 fission and cluster radioactivity \cite{ishwar}.
 The momentum-dependence \emph{V$_{mdi}^{ij}$} of the nn interactions, which may optionally be used in IQMD, is fitted to the experimental data
 in the real part of the nucleon optical potential.

 \section{Results and discussion}

  For the present study, we simulate several thousands events of each
  reaction at incident energies around E$_{bal}$ in small steps of 10 MeV/nucleon.
   In particular, we simulate the reactions
$^{24}$Mg+$^{24}$Mg, $^{58}$Cu+$^{58}$Cu, $^{72}$Kr+$^{72}$Kr,
$^{96}$Cd+$^{96}$Cd, $^{120}$Nd+$^{120}$Nd, $^{135}$Ho+$^{135}$Ho,
having N/Z = 1.0 and reactions $^{24}$Ne+$^{24}$Ne,
$^{58}$Cr+$^{58}$Cr, $^{72}$Zn+$^{72}$Zn, $^{96}$Zr+$^{96}$Zr,
$^{120}$Sn+$^{120}$Sn, and $^{135}$Ba+$^{135}$Ba, having N/Z =
1.4, respectively,
 in the whole range of colliding geometry. The
colliding geometry is divided into four impact parameter bins of
0.15 $<$ $\hat{b}$ $<$ 0.25 (BIN 1),
 0.35 $<$ $\hat{b}$ $< $0.45 (BIN 2), 0.55 $<$ $\hat{b}$ $<$ 0.65 (BIN 3),
and 0.75 $<$ $\hat{b}$ $<$ 0.85 (BIN 4), where $\hat{b}$ =
b/b$_{max}$.

\begin{figure}[!t] \centering \vskip -0.2cm
\includegraphics[angle=0,width=12cm]{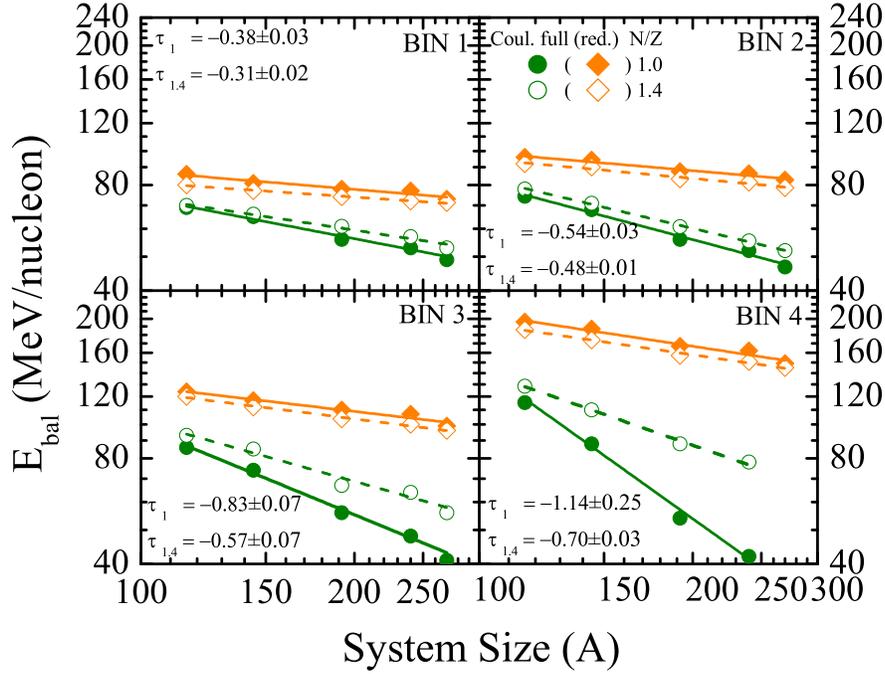}
\vskip 0.5cm \caption{{(Color online)} E$_{bal}$ as a function of
combined mass of system for different impact parameter bins. Solid
(open) symbols are for systems having N/Z = 1.0 (1.4). The solid
(dashed) lines are power law fit $\varpropto A^{\tau}$. $\tau$
values for full Coulomb calculations are displayed in figure. The
values of $\tau$ for reduced Coulomb calculations are given in the
text.}\label{fig1}
\end{figure}

 Here N/Z is changed by keeping the mass fixed. We use
anisotropic standard isospin- and energy-dependent nn cross
section $\sigma$ = 0.8 $\sigma$$_{NN}$$^{free}$. The details about
the elastic and inelastic cross sections for proton-proton and
proton-neutron collisions can be found in Ref. \cite{hart98}. The
cross sections for neutron-neutron collisions are assumed to be
equal to the proton-proton cross sections. We also use soft
equation of state along with momentum-dependent interactions
(MDI). The results with the above choice of equation of state and
cross section were in good agreement with the data. The reactions
are followed until the transverse flow saturates. The saturation
time varies form 100 fm/c for lighter masses to 300 fm/c for
heavier masses. For transverse flow, we use the quantity
"\textit{directed transverse momentum $\langle
p_{x}^{dir}\rangle$}" which is defined as \cite{sood04}
\begin {equation}
\langle{p_{x}^{dir}}\rangle = \frac{1} {A}\sum_{i=1}^{A}{sign\{
{y(i)}\} p_{x}(i)},
\end {equation}
where $y(i)$ is the rapidity and $p_{x}$(i) is the momentum of
$i^{th}$ particle. The rapidity is defined as
\begin {equation}
Y(i)= \frac{1}{2}\ln\frac{{\textbf{{E}}}(i)+{{\textbf{p}}}_{z}(i)}
{{{\textbf{E}}}(i)-{{\textbf{p}}}_{z}(i)},
\end {equation}
where ${\textbf{E}}(i)$ and ${\textbf{p}_{z}}(i)$ are,
respectively, the energy and longitudinal momentum of $i^{th}$
particle. In this definition, all the rapidity bins are taken into
account. A straight line interpolation is used to calculate
E$_{bal}$.

\par
\begin{figure}[!t] \centering
\vskip 0.15cm
\includegraphics[angle=0,width=10cm]{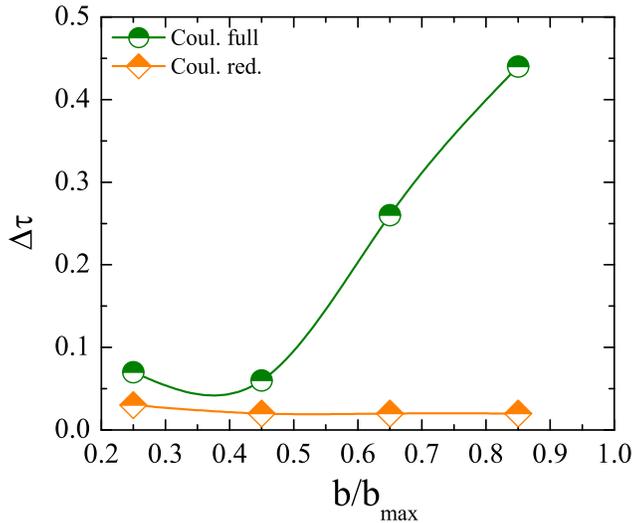}
\vskip 0.5cm \caption{{(Color online)} $\tau$ as a function of
reduced impact parameter. Lines are only to guide the
eye.}\label{fig3}
\end{figure}
 \par
  Figure 1
displays the mass dependence of E$_{bal}$ for four impact
parameter bins. The solid (open) green circles indicate
 E$_{bal}$ for systems with lower (higher) N/Z. Lines are power law fit $\varpropto$ A$^{\tau}$. E$_{bal}$ follows a power law behavior
$\varpropto$ A$^{\tau}$ for both N/Z = 1 and 1.4 ($\tau$ being
labeled as $\tau_{1.0}$ and $\tau_{1.4}$ for systems having N/Z =
1 and 1.4, respectively) at all colliding geometries. Also, for
the clarity, we did not include the very light system (A=48) as
isospin effects are negligible at all colliding geometries for
this system. \par
From fig. 1 we see that isospin effects are
clearly visible for all the
  four bins as neutron-rich system has higher E$_{bal}$ throughout the mass range in agreement with the previous
  studies \cite{pak97,li96,gaum10}.
\begin{figure}[!t] \centering
\vskip -0.1cm
\includegraphics[angle=0,width=10cm]{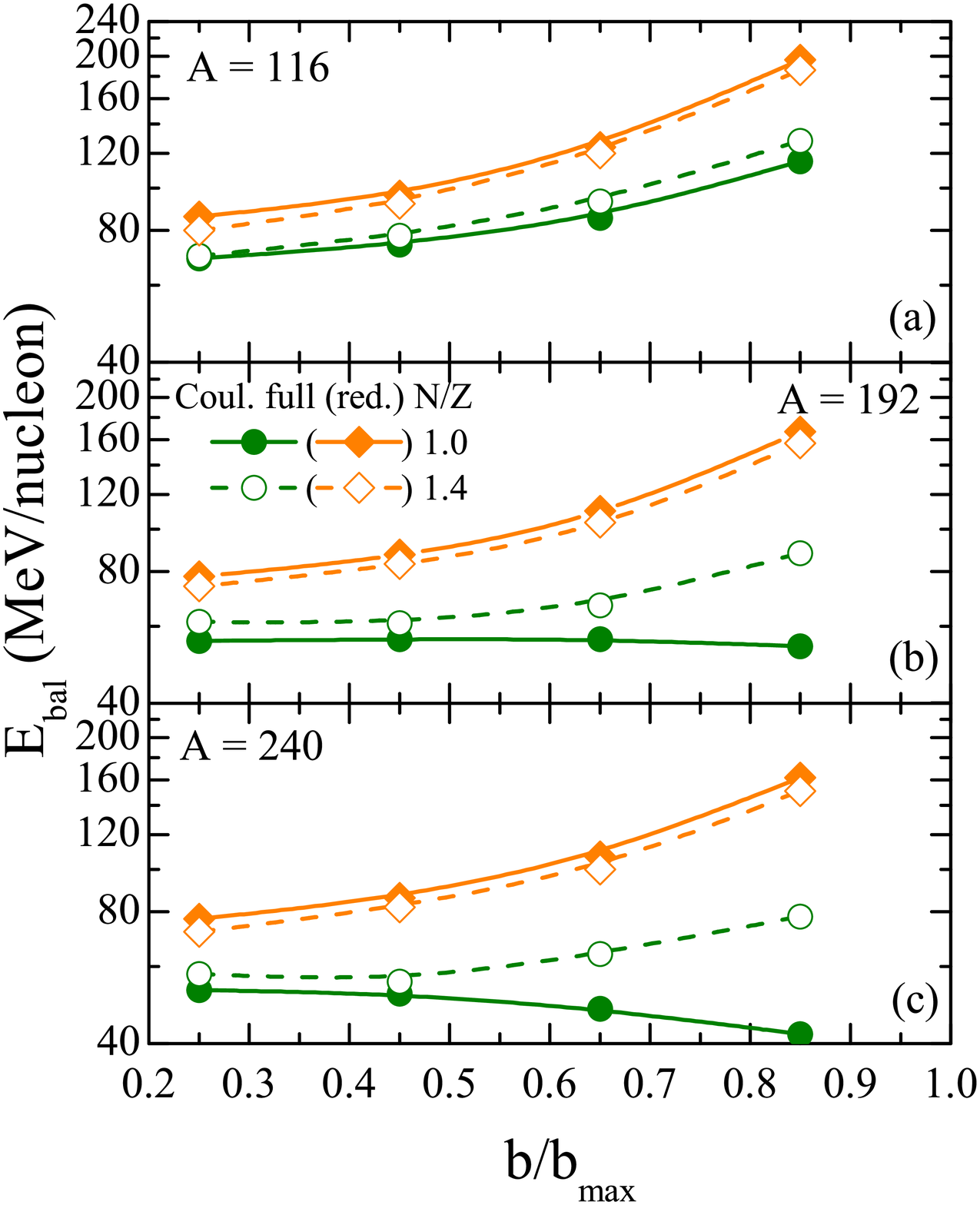}
\vskip 0.65cm \caption{{(Color online)} E$_{bal}$ as a function of
impact parameter for different system masses. Symbols have same
meaning as in Fig.1. Lines are only to guide the eye.
\cite{gaum210}}\label{fig4}
\end{figure}
The magnitude of the isospin effects increases with the increase
in the mass of the system at all colliding geometries. The effect
is much more pronounced at larger colliding geometries. One can
see that the difference between $\tau_{1.0}$ and $\tau_{1.4}$
increases with increase in the impact parameter. In Ref.
\cite{gaum210}, one of us and coworkers studied the isospin
effects on the mass dependence of E$_{bal}$ for BIN 2. There the
Coulomb potential was reduced by a factor of 100 and was showed
that the Coulomb repulsion plays dominant role over symmetry
energy in isospin effects on E$_{bal}$ as well as its mass
dependence
 at semi central colliding geometry (BIN 2). Since here we plan to extend
that study over a full range of colliding geometry, so here also
we reduce the Coulomb potential by a factor of 100 and calculate
the E$_{bal}$ throughout the mass range at all colliding
geometries. Solid (open) diamonds
 represent E$_{bal}$ calculated with reduced Coulomb for systems with lower (higher) neutron content.
  Lines are power law fit $\varpropto$ A$^{\tau}$. The values of $\tau_{1.0}$ ($\tau_{1.4}$) are
-0.17 (-0.14), -0.17 (-0.19), -0.24 (-0.26),
 and -0.31 (-0.29)
for BIN 1, BIN 2, BIN 3, and BIN 4, respectively. Interestingly,
we find that the magnitude of isospin effects (difference in
E$_{bal}$ for a given pair) is now nearly same throughout the mass
range which indicates that the effect of symmetry energy is
uniform throughout the mass range. This is true for all the
colliding geometries. This is supported by
 Ref. \cite{sood04} where Sood and Puri studied the average density as a function of mass of the system
 (throughout the periodic table) at incident energies equal to E$_{bal}$ for each
given mass. There they found that although both E$_{bal}$ and
average density follows a power law behavior $\varpropto$
A$^{\tau}$, E$_{bal}$ decreases more sharply with the
 combined mass of the system (with $\tau$ = -0.42), whereas the average
density (calculated at incident energy equal to E$_{bal}$) is
almost independent of the mass of the system with $\tau$ = -0.05.
It is worth mentioning here that the trend will be different at
fixed incident energy in which case density increases with
increase
 in the mass of the system \cite{blatt,khoa}. From fig. 1 one can also see that the enhancement in E$_{bal}$ (by reducing
Coulomb) is more in heavier
 systems as compared to lighter systems for
 all colliding geometries. The effect is more pronounced at higher colliding geometries. Moreover, throughout the mass range at all
  colliding geometries, the neutron-rich systems
 have less E$_{bal}$ as compared to neutron-deficient systems when we reduce the Coulomb. This trend is quite the opposite to the one
 which we have when we have full Coulomb. This (as explained in Ref. \cite{gaum210} also) is due to the fact that the reduced
Coulomb repulsion leads to higher E$_{bal}$. As a result, the
density achieved during the course of the reaction will be more
due to which the impact of the repulsive symmetry energy will be
more in neutron-rich systems, which in turn leads to less
E$_{bal}$ for neutron-rich systems.
\par
\begin{figure}[!t] \centering
\vskip 0.5cm
\includegraphics[angle=0,width=10cm]{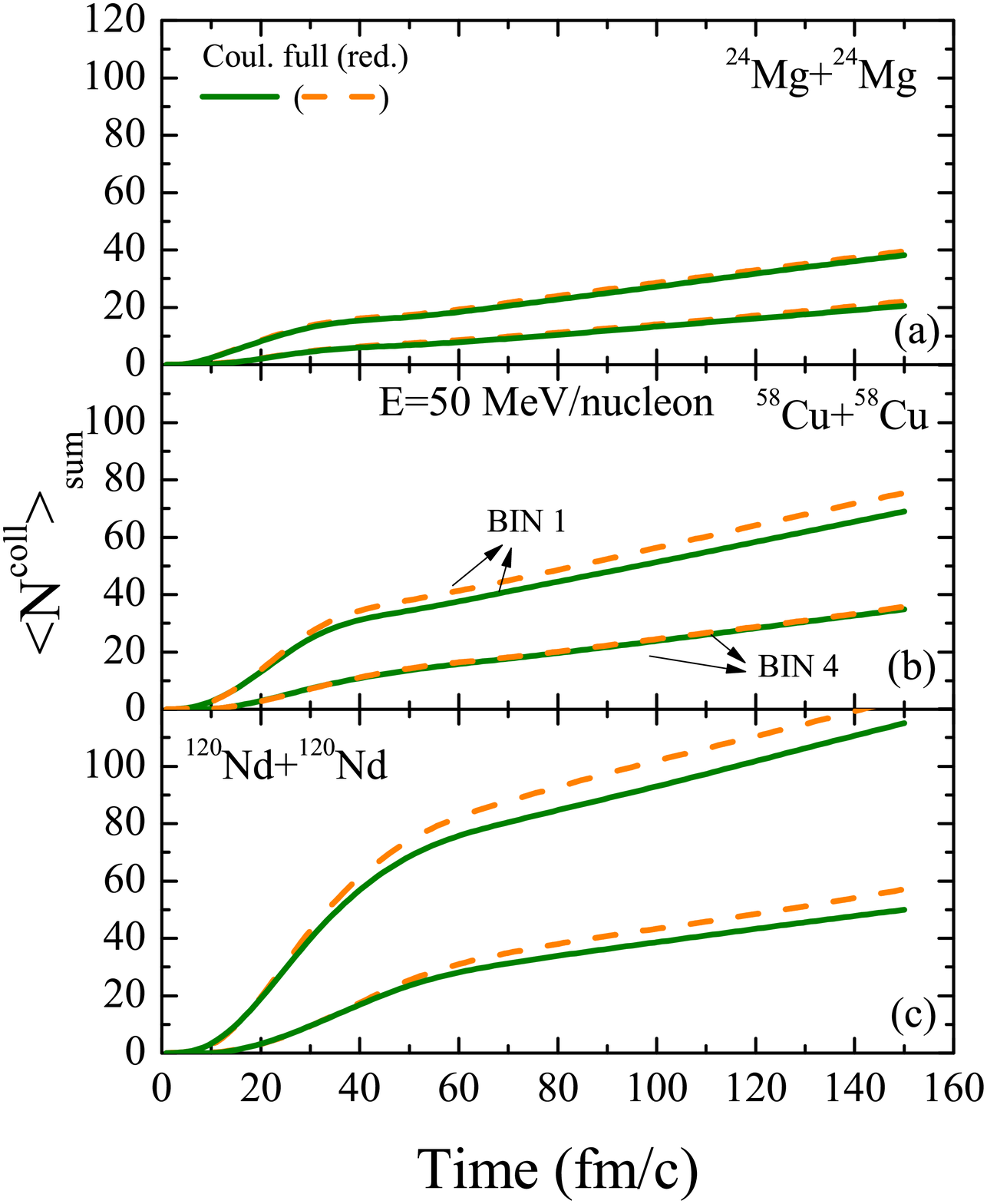}
\vskip 0.5cm \caption{(Color online) The time evolution of number
of collisions for various system masses at 50 MeV/nucleon for BIN
1 and BIN 4. Various lines are explained in the text
.}\label{fig4}
\end{figure}

In fig. 2, we show the variation in $\Delta \tau$ as a function of
colliding geometry for both full and reduced Coulomb where $\Delta
\tau = \tau_{1.4} - \tau_{1.0}$. Half filled circles (diamonds)
represent calculations with full (reduced) Coulomb. $\Delta \tau$
values are plotted at upper limit of impact parameter for each
bin. Lines are only to guide the eye.
 From figure (green circles), we see that the difference between $\tau_{1.4}$ and $\tau_{1.0}$ increases with increase in impact parameter which is
 due to the sharp increase in value of $|\tau_{1}|$, whereas when we reduce the Coulomb (orange diamonds), this difference is negligible, which shows an
  enhanced effect of Coulomb at higher impact parameters.
 \par
 In fig. 3a, 3b, and 3c, we display E$_{bal}$ as a function of $\hat{b}$ for masses 116, 192, and 240, respectively, for both full
 and reduced Coulomb. Symbols have the same meaning as in fig. 1. For full Coulomb (green circles), for all the masses at all colliding
 geometries, system with higher N/Z has larger E$_{bal}$ in agreement with previous studies \cite{li96,pak97}. Moreover,
  the difference between E$_{bal}$ for a given mass
 pair, increases with increase in colliding geometry. This is more clearly visible in heavier masses. Also for N/Z = 1.4, E$_{bal}$ increases with increase in
 impact parameter as expected \cite{chugh10,mag00}. However for N/Z = 1, this is true only for lighter mass system such as A = 116. For heavier masses E$_{bal}$ infact
 begins to decrease with increase in impact parameter in contrast to the previous studies \cite{li96,pak97,gaum10,chugh10,mag00}. However, when we reduce the Coulomb (by a factor of 100 (diamonds)),
 we find that:
 \par
  (i) Neutron-rich systems have smaller E$_{bal}$ as compared to neutron-deficient systems as mentioned
  previously also. This is true at all the
colliding geometries throughout the mass range. This clearly shows
the dominance of Coulomb repulsion over symmetry energy in isospin
effects throughout the mass range at all colliding geometries.
\par
(ii) The difference between E$_{bal}$ for systems with different
N/Z remains almost constant as a function of colliding geometry
which indicates that the effect of symmetry energy is uniform
throughout the range of $\hat{b}$ as well. This also shows that
the large differences in E$_{bal}$ values for a given isobaric
pair are due to the Coulomb repulsions.
 \par
 To see the role of Coulomb repulsion on nn collisions, in fig. 4
 we display no. of collisions for A = 48, 116, and 240 for
 calculations with and without Coulomb potential at central (BIN
 1) and peripheral (BIN 4) colliding geometry at incident energy of 50 MeV/nucleon. Solid (dashed)
 lines represent Coulomb full (reduced) calculations. Higher (lower)
 lines represent the results for BIN 1 (BIN 4). We find that
 Coulomb decreases the no. of collisions in medium and heavier
 mass systems for central bin whereas for peripheral bin, the
 effect of Coulomb on collisions is only for heavier masses (lower
 panel).

\section{Summary}
We have studied the isospin effects in the disappearance of flow
as well as its mass dependence throughout the mass range for two
sets of isobaric systems with N/Z = 1 and 1.4 in the whole range
of colliding geometry. Our results clearly demonstrate the
dominance
 of Coulomb repulsion in isospin effects on E$_{bal}$ as well as its
mass dependence throughout the range of colliding geometry. The
above study also shows that the effect of symmetry energy on
E$_{bal}$ is uniform throughout the mass range and colliding
geometry.
\par
This work has been supported by a grant from Indo-French Centre
For The Promotion Of Advanced Research (IFCPAR) under project no.
4104-1.

\end{document}